# Observation of Hybrid-Order Topological Pump in a Kekulé-Textured Graphene Lattice

Tianzhi Xia,[1] Yuzeng Li,[*,1] Qicheng Zhang,[1] Xiying Fan,[1] Meng Xiao,[1,2] and Chunyin Qiu[*,1]

[1]School of Physics and Technology, Wuhan University, Wuhan 430072, China
[2]Wuhan Institute of Quantum Technology, Wuhan 430206, China

[*]To whom correspondence should be addressed: yuzengli@whu.edu.cn; cyqiu@whu.edu.cn

*Abstract.* Thouless charge pumping protocol provides an effective route for realizing topological particle transport. To date, the first-order and higher-order topological pumps, exhibiting transitions of edge-bulk-edge and corner-bulk-corner states, respectively, are observed in a variety of experimental platforms. Here, we propose a concept of hybrid-order topological pump, which involves a transition of bulk, edge, and corner states simultaneously. More specifically, we consider a Kekulé-textured graphene lattice that features a tunable phase parameter. The finite sample of zigzag boundaries, where the corner configuration is abnormal and inaccessible by repeating unit cells, hosts topological responses at both the edges and corners. The former is protected by a nonzero winding number, while the latter can be explained by a nontrivial vector Chern number. Using our skillful acoustic experiments, we verify those nontrivial boundary landmarks and visualize the consequent hybrid-order topological pump process directly. This work deepens our understanding to higher-order topological phases and broadens the scope of topological pumps.

*Introduction.* Topology, originally a branch of mathematics, has become an important concept in different fields of physics [1-6]. Thouless pump provides one of the simplest manifestations to understand the band topology in quantum systems. It was first proposed by Thouless when he studied the particle transport in one-dimensional (1D) periodic structures with adiabatic time evolution. He found out that the dynamic edge-bulk-edge pumping shares the same topological origin as the static two-dimensional (2D) Chern insulator [7,8], where one of the momentum coordinates is replaced by an adiabatically varied cyclic parameter [Fig. 1(a)]. As such, this dynamic pumping process can be dictated by a nontrivial integer topological invariant (i.e., first Chern number). Similar theory has also been extended to explore four-dimensional (4D) quantum Hall effect characterized by a 4D topological invariant (known as second Chern number) [9,10]. In this case, a 2D Thouless pump is implemented to a 2D spatial system, resorting to two additional external parameters [10]. With the discovery of higher-order topological insulators [11-15], referring to $d$-dimensional topological systems hosting nontrivial $(d-N)$-dimensional boundary states ($N>1$), higher-order topological pumps have been proposed to connect the quantized corner to corner charge flow with the unconventional bulk-boundary correspondence [16-21], as illustrated in Fig. 1(b). To date, both the conventional (or first-order) and higher-order Thouless pumps have been realized in diverse experimental platforms, from cold atom systems to various artificial crystals of classical waves [22-32].

Note that both the first-order and higher-order pumping procedures involve only a simple transition between the bulk and edge (or corner) states. To the best of our knowledge, so far there is no realization of topological pump involving the states of more spatial dimensions simultaneously, e.g., the bulk, edge, and corner states in a 2D system. To achieve such new kind of topological pumps, dubbed hybrid-order pumps latter, a suitable band gap formed *between* the 2D bulk and 1D edge bands should be designed for tolerating a continuous pumping of the zero-dimensional (0D) corner states [Fig. 1(c)]. This is much more difficult than that demanded for a higher-order pump: the validity of the higher-order topological invariant relies on a concrete selection of the periodical unit cell [11-15], which makes the boundary morphology and thus the edge modes predetermined by the corner configuration in the finite system formed by repeating unit cells.

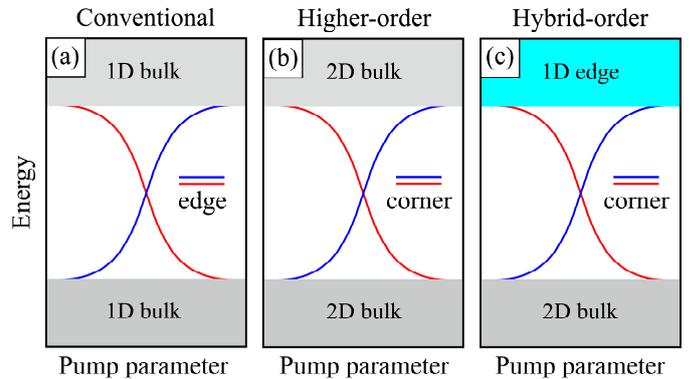

FIG. 1. Schematic illustrations for three distinct topological pumps. (a) Conventional (first-order) pump in 1D systems. (b) Higher-order topological pump in 2D systems. (c) Hybrid-order topological pump in 2D systems. The red and blue lines sketch 0D states confined to different edges (or corners). In contrast to the conventional and higher-order pumps that exhibit edge-bulk-edge and corner-bulk-corner transports within a parameter cycle, the hybrid-order pump exhibits a transition among the bulk, edge, and corner states.

In this work, we report an experimental observation of hybrid-order topological pump in acoustic systems. As shown in Fig. 2(a), we consider a 2D Kekulé-textured graphene (KTG) lattice featuring three inequivalent hoppings, where the variable phase ($\phi$) acts as the cyclic pumping parameter. Particularly, we focus on the obtuse angle corners intersected by two zigzag boundaries. In this unusual boundary configuration, which is inaccessible by repeating any minimal unit cells, the system exhibits corner states that traverse the edge and bulk states continuously with the phase parameter. Using acoustic cavity-tube structures, we experimentally observe the highly appealing hybrid-order topological pumping process, accompanying convincing experimental evidence for the topological responses at the 1D edges and 0D corners.

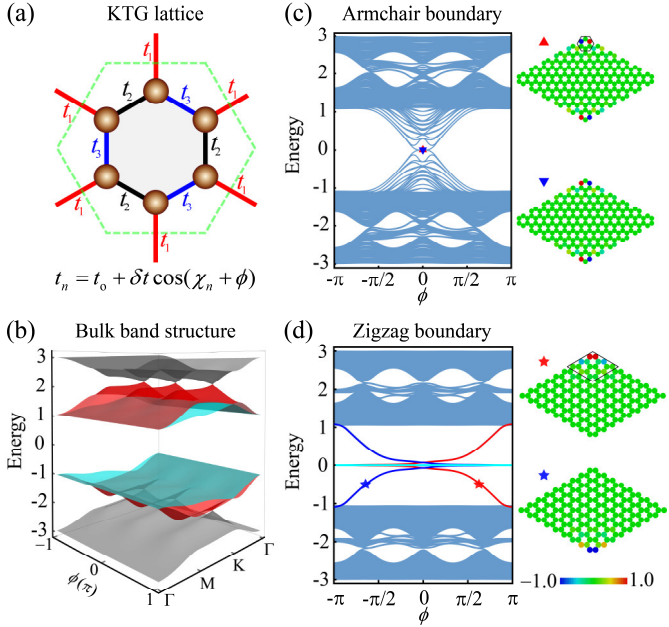

FIG. 2. Tight-binding model. (a) Unit cell of the KTG lattice, which features textured hoppings $t_1$, $t_2$, and $t_3$. (b) Evolution of the band structure as the phase parameter $\phi$, evaluated with fixed $t_o = -1$ and $\delta t = -0.7$. (c) Left: $\phi$-dependent energy spectrum for a finite system with armchair boundary. Right: Eigenfields for the degenerate zero-energy corner states highlighted in the spectrum. (d) Similar to (c), but for the system with zigzag boundary. Note that the corner configuration cannot be formed by simply repeating KTG unit cells or any other minimal unit cells of 6 orbitals, but can be achieved by duplicating the rhombus supercell of 18 orbitals. The coexistence of the nontrivial edge (cyan) and corner (red and blue) modes offers an opportunity to realize the hybrid-order topological pumping in the zigzag-boundary KTG lattice.

*Tight-binding model.* As shown in Fig. 2(a), we start with a 2D KTG model that is distorted from the graphene lattice of uniform hopping $t_o$ [33,34]. The KTG lattice features three inequivalent nearest-neighboring hoppings $t_n = t_o + \delta t \cos(\chi_n + \phi)$, where $\delta t$ characterizes the modulation strength of the hoppings, and $\phi \in [-\pi, \pi]$ is a tunable phase parameter under the specific bias $\chi_n = 2(n-1)\pi/3$ ($n = 1 \sim 3$). Note that the lattice constant $a$ of the honeycomb lattice becomes $\sqrt{3}a$ for the KTG lattice. In addition to time-reversal symmetry, the system of a general $\phi$ has sublattice symmetry (or chiral symmetry). It is of particular interest that, as exhibited in Fig. 2(b), the band structure of the KTG lattice always hosts a wide spectral gap centered at the zero energy, besides the mini gaps appearing repeatedly between the highest (lowest) two bands.

Now we consider the topological boundary manifestations in finite-sized samples. Armchair and zigzag boundaries are two typical boundary truncations for general honeycomb lattices. In our KTG lattice, however, unlike the armchair boundary accessible by simply duplicating unit cells, the zigzag boundary cannot be achieved without destroying the integrity of the 6-orbital KTG unit cell. Instead, it can be realized by repeating rhombus supercells of 18 orbitals. Figure 2(c) presents the spectral flow against the phase parameter $\phi$ for the case of natural armchair boundary. It shows that near $\phi = 0$, two degenerate zero-energy states emerge at the corners of the obtuse angle, either symmetric or antisymmetric about the center of the finite diamond-shaped structure. The corner states can be explained by a $Z_2$ topological invariant or a topological index defined at the high symmetry points in momentum space [34-37]. For the system with unusual zigzag boundary, however, novel phenomena emerge in the primary bulk gap. As shown in Fig. 2(d), highly-degenerate flat bands (cyan curves) appear at zero energy and divide the primary gap into two pieces. They are edge states inherited from the original graphene lattice of zigzag boundary, which can be characterized by a nontrivial winding number [38,39] (see *Supplemental Materials*). More intriguingly, there are two sets of corner states spanning the two separate primary gaps with the continuum evolution of $\phi$, each of which emerges pairwise in energy at the same corner due to the chiral symmetry. Besides, the corner states associated to the samples of $\pm\phi$ are localized in opposite obtuse angle corners, as exemplified by the field distributions for $\phi = \pm 5\pi/8$. Therefore, for any of the two divided primary gaps, one can imagine a hybrid-order topological pumping process that involves the bulk, edge, and corner states simultaneously.

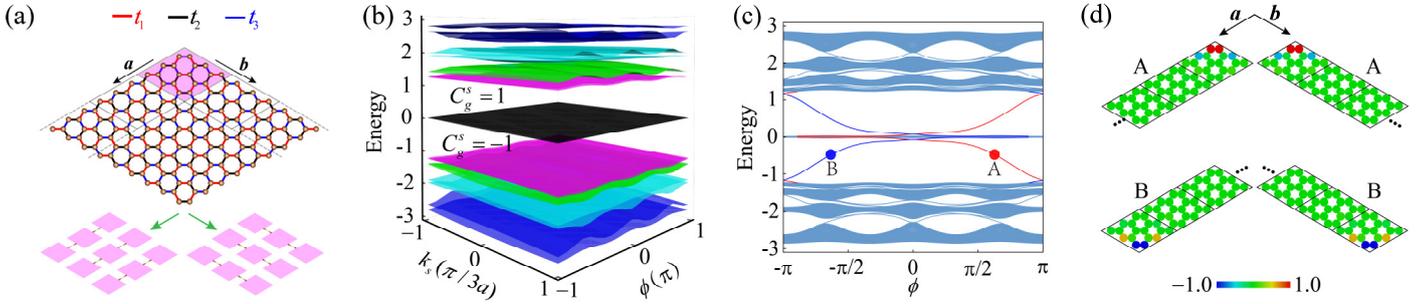

FIG. 3. Physical origin of the gapless corner states. (a) A zigzag-boundary KTG lattice formed by duplicating rhombus supercells (shadowed). It can be decoupled into supercell chains along the **a** and **b** directions by removing the inter-supercell hoppings in the **b** and **a** directions, respectively. (b) Global evolution of the band structure for the 1D periodic chain in the **a**(**b**) direction, which emulates the bulk dispersion of a 2D Chern insulator under the assistance of the effective momentum $\phi$. Here $C_g^s$ highlights the Chern number for the two primary gaps near the zero-energy flat bands. (c) $\phi$-dependent energy spectrum for the **a**(**b**)-directed chain of finite length, which exhibits in-gap edge modes (blue and red lines) protected by the gap Chern numbers. (d) Field distributions of the edge modes labeled in (c), which together form the top (A) and bottom (B) corner modes in Fig. 2(d).

*Topological nature of the gapless corner states*. Having known that the edge states are protected by a nontrivial winding number, below we explain the physical origin for the gapless corner states in the zigzag-boundary systems, inspired by the top-to-bottom scheme introduced in Ref. [40], from which one constructs 2D corner states from two independent 1D systems with 0D edge states. To do this, we consider a finite system of $N \times N$ rhombus supercells, and decompose it into $N$ 1D supercell chains in the **a** and **b** directions [Fig. 3(a)]. Each supercell chain constitutes a 2D Chern insulator when introducing the phase parameter $\phi$ as an effective momentum for the second, synthetic dimension. Figure 3(b) shows the bulk band structure for the synthetic 2D Chern insulator, where the chiral symmetry-related bands are plotted with the same color for clarity. In particular, the two nearly degenerate flat bands (black) share a similar physical origin to those edge modes exhibited in Fig. 2(d), while the states are now localized at the *supercell's* zigzag boundaries. According to the bulk-boundary correspondence, the topological edge states of the synthetic Chern insulator are related to its gap Chern number, which can be defined as

$$C_g^s = \frac{1}{2\pi i} \int_{BZ} \mathrm{Tr}[F(k_s, \phi)] \, dk_s d\phi. \quad (1)$$

Here, $s = \boldsymbol{a}, \boldsymbol{b}$ and $F(k_s, \phi)$ represents non-Abelian Berry curvature calculated for all bands below the target gap [40,41]. Specifically, the gap Chern number $C_g^s = 1$ (or $-1$) for the primary gap above (below) the flat bands. This suggests that each gap hosts one gapless edge band when the synthetic Chern insulator is truncated in the **a**(**b**) direction. This can be seen in Fig. 3(c), the projected energy spectrum along the synthetic momentum $\phi$. Figure 3(d) exemplifies the edge state distributions for $\phi = \pm 5\pi/8$ in the lower primary gap. When both the **a**- and **b**-directed supercell chains host nontrivial edge states decaying exponentially away from the ends, corner states are concluded in the 2D zigzag-boundary sample whose couplings can be decomposed into two independent components in these directions [40,41]. (More specifically, the top edge localizations in $\phi = 5\pi/8$ contribute the top corner states, while the bottom edge localizations for $\phi =$ $-5\pi/8$ contribute the bottom corner states [Fig. 2(d)].) Therefore, the corner states of the original zigzag-boundary KTG lattice can be characterized by the two non-zero gap Chern numbers $C_g^a$ and $C_g^b$, or written together as a vector one, $\boldsymbol{C} = (C_g^a, C_g^b)$. More specifically, the vector Chern numbers $\boldsymbol{C}_1 = (-1, -1)$ and $\boldsymbol{C}_2 = (1,1)$ explain the corner states of the primary gaps below and above the zero energy, respectively. Ultimately, the coexistence of the topological corner and edge states in the diamond-shaped sample, protected respectively by the nontrivial vector Chern numbers and winding numbers, enables a unique hybrid-order pumping process when considering a continuous evolution of the phase parameter $\phi$.

*Acoustic realization of the tight-binding model.* The above tight-binding model can be emulated precisely by our acoustic system, where the orbitals and hoppings are mimicked by air-filled cavity resonators and narrow tubes [42-45], respectively. As shown in Fig. 4(a), each unit cell of our acoustic KTG lattice includes six identical hexagonal prism cavities of side length $l = 5.0$ mm and height $H = 32.8$ mm. The latter results in a dipole resonant mode of 5.27 kHz, which is far from the other resonant modes. (In our simulations, the sound speed is set as 346 m/s.) Every two nearest acoustic resonators are connected by two rectangular tubes of constant aspect ratio $1: 3$. Providing that the couplings are approximately proportional to the cross-sectional areas of the narrow tubes, as shown in Fig. 4(b), we realize the desired couplings by modulating the cross-sectional areas $S_n = S_o + \delta S \cos(\chi_n + \phi)$, with $S_o = 20.3$ mm$^2$ and $\delta S = 12.0$ mm$^2$. The resultant effective hoppings (colored dots) capture well the cosine line shape $t_n = t_o + \delta t \cos(\chi_n + \phi)$, associated with $t = -424.5$ Hz and $\delta t = -216.7$ Hz. To demonstrate the topological response of the system, Fig. 4(c) exemplifies the eigenvalue spectrum with $\phi = 5\pi/8$, simulated for a sample of $3 \times 3$ supercells. In addition to the midgap edge modes, the energy spectrum features two chiral symmetry-related 0D modes at the top corner of the sample, as visualized further by their acoustic field distributions (see insets).

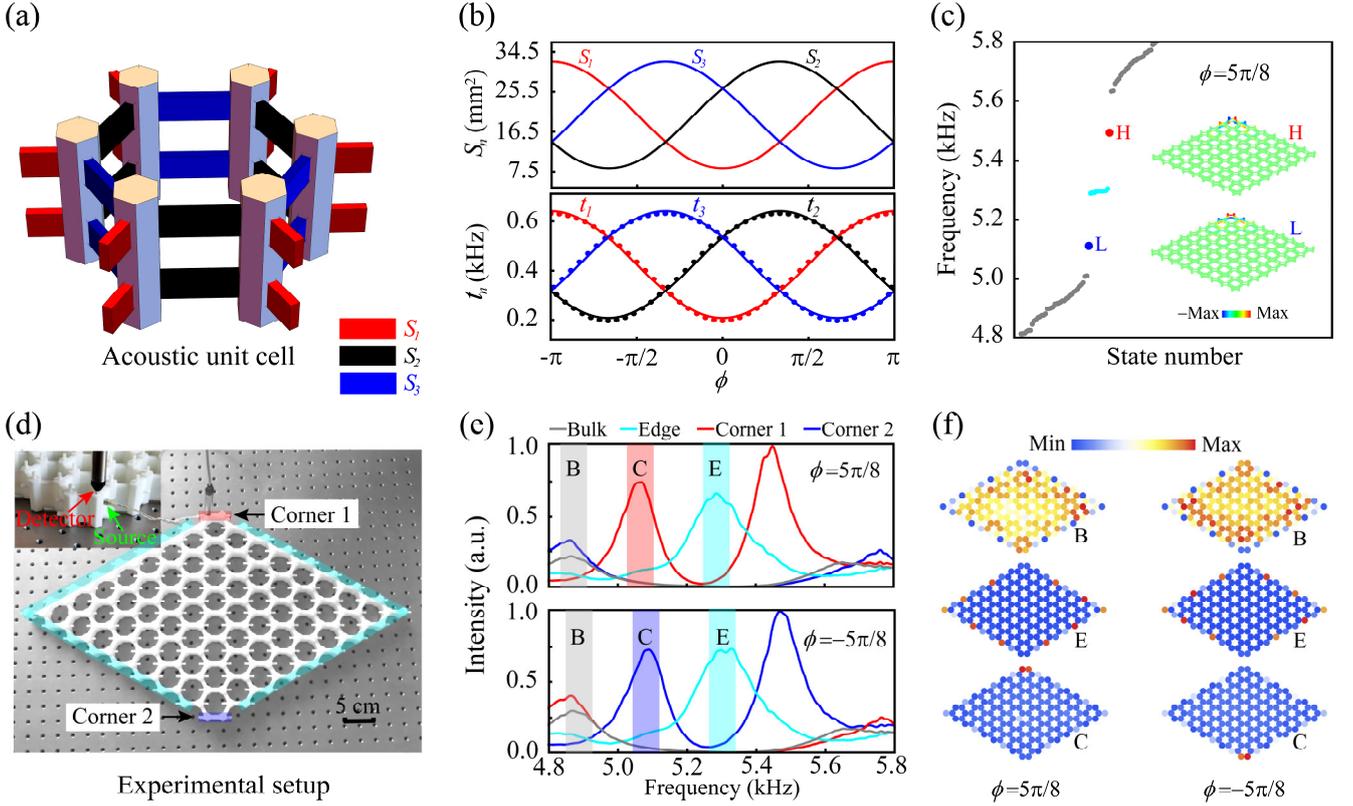

FIG. 4. Acoustic emulation of the tight-binding model and experimental evidence for the topological states. (a) Unit cell geometry of our acoustic KTG lattice, which features three types of coupling tubes of different cross-sectional areas, $S_1$, $S_2$, and $S_3$. (b) Top: $\phi$-dependent cross-sectional areas designed for the three coupling tubes. Bottom: Associated effective couplings (color dots), which are fitted well by cosine functions (color lines). (c) Eigenfrequency spectra simulated for a finite sample with $\phi = 5\pi/8$. Insets: Pressure field distributions extracted for the two chiral symmetry-related corner states. (d) A photograph of the zigzag-boundary sample, where the corner and edge sites are colored for clarity. The inset shows an enlarged view around the corner 1. (e) Pressure intensity spectra averaged for the bulk, edge, and corner sites. The results are exemplified with the samples of $\phi = \pm 5\pi/8$. (f) Intensity patterns extracted for the bulk (B), edge (E), and corner (C) states, respectively, which are averaged over the frequency windows shadowed in (e).

Below we present our experimental evidence for the topological edge and corner states. Here, we focus on the sample with $\phi = 5\pi/8$ first. As shown in Fig. 4(d), our sample fabricated by 3D-printing consists of $3 \times 3$ supercells, i.e., 162 acoustic cavity resonators in total. To implement local measurements, two small holes are perforated in each cavity for inserting the sound source and probe, which are sealed when not in use. Both the input and output signals are recorded and frequency-resolved with a multi-analyzer system (B&K Type 4182). The samples are divided into non-overlapping spatial domains to extract the information of the bulk, edge, and corner states. The top panel of Fig. 4(e) presents the site-averaged bulk, edge and corner spectra. Clearly, the bulk spectrum (gray curve) shows a wide band gap centered at 5.27 kHz (associated to the zero energy in tight-binding model), around which the edge spectrum (cyan curve) shows a predominant peak. By contrast, a pair of chiral symmetry-related peaks appear in the spectrum of the corner 1 (red curve), which are missed in the spectrum of the corner 2 (blue curve), as predicted in Fig. 2(d). Similar phenomena can be observed in the case of $\phi = -5\pi/8$ [Fig. 4(e), bottom panel], where the corner states appear in the corner 2 (blue curve). To further identify the bulk, edge, and corner states, we have integrated the pressure intensities over several typical frequency ranges for each site individually. As shown in Fig. 4(f), the site-resolved acoustic patterns exhibit clear spatial characteristics of the bulk, edge, and corner states as expected. It is worth pointing out that in graphene-like lattices, although extensively unveiled in armchair-boundary configurations, the higher-order corner states have not been observed so far in any system with zigzag boundaries.

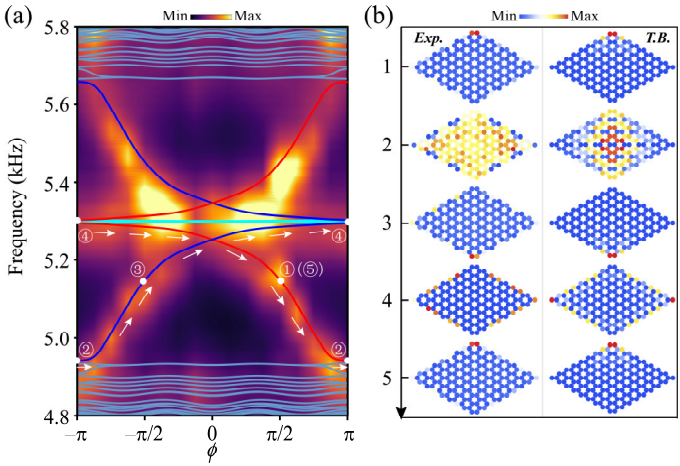

FIG. 5. Experimental observation of the hybrid-order topological pump. (a) Global evolution of the bulk, edge, and corner spectra (color scale), together with those tight-binding predications (color lines) for comparison. The white arrows highlight an intact pumping process. (b) Field distributions scanned for the states labeled from ① to ⑤, comparing with those tight-binding results. The data exhibit a clear hybrid-order topological pump that involves the corner, bulk, and edge states simultaneously.

*Observation of the hybrid-order topological pump.* To reflect the global evolution of the bulk, edge, and corner states within one pumping cycle $\phi \in [-\pi, \pi]$, we fabricated 17 diamond-shaped samples at the $\phi$-step of $\pi/8$, and measured their topological manifestations individually as before. All experimental data match well our full-wave simulated bulk, edge, and corner spectra (see [Supplemental Materials](#)). For conciseness, here we count the bulk, edge, and corner spectra together and present the superimposed data (color scale) in Fig. 5(a). It exhibits an excellent agreement with our theoretical prediction (color lines), especially for the corner modes across the two primary band gaps around 5.3 kHz. The band broadening stems mostly from the unavoidable viscous and thermal dissipations inside the acoustic structure. More importantly, the spectral flow unveils a hybrid-order topological pumping process where the corner states traverse the bulk and edge states smoothly, as guided by the white arrows in the lower primary gap. To directly visualize the intricate hybrid-order topological pump, we display the site-resolved intensity patterns measured for a set of typical configurations [Fig. 5(b), left panel]. Clearly, the acoustic state evolves from the top corner ① to the bulk ②, consequently to the bottom corner ③, and then it goes back to the initial state ⑤ through the edge ④, during which the band gap is never closed. This reproduces the pumping process predicted by the tight-binding model [Fig. 5(b), right panel].

*Conclusions.* By fully exploiting the controllability of acoustic metamaterials, we have designed and fabricated a series of acoustic KTG lattices to realize the novel hybrid-order topological pumping protocol. Tracking the global evolution of the acoustic states with the cyclic parameter, our experimental results exhibit an intact corner-bulk-corner-edge-corner transition, which involves the 0D corner, 1D edge, and 2D bulk states simultaneously. All experimental data agree well with the theoretical predictions. One may expect that both the corner states and edge states involved here would be considerably robust against disorders and defects, given the protection by the gap Chern number and winding number. Our findings open a new path for unveiling more complex topological pumping physics.

**Acknowledgements**. This project is supported by the National Natural Science Foundation of China (Grant No. 11890701, 12374418, 12004287, and 12104346), the Young Top-Notch Talent for Ten Thousand Talent Program (2019-2022).

**References:**
[1] K. V. Klitzing, The quantized Hall effect, Rev. Mod. Phys. **58**, 519 (1986).
[2] C. L. Kane and Liang Fu, Time reversal polarization and a Z2 adiabatic spin pump, Phys. Rev. B **74**, 195312 (2006).
[3] X. Qi, T. L. Hughes, and S. Zhang, Topological field theory of time-reversal invariant insulators, Phys. Rev. B **78**, 195424 (2008).
[4] X. Qi, T. L. Hughes, S. Raghu, and S. Zhang, Time-reversal-invariant topological superconductors and superfluids in two and three dimensions, Phys. Rev. Lett. **102**, 187001 (2009).
[5] M. Z. Hasan and C. L. Kane, Colloquium: Topological insulators, Rev. Mod. Phys. **82**, 3045 (2010).
[6] X. Qi and S. Zhang, Topological insulators and superconductors, Rev. Mod. Phys. **83**, 1057 (2011).
[7] D. J. Thouless, Quantization of particle transport, Phys. Rev. B **27**, 6083 (1983).
[8] R. Citro and M. Aidelsburger, Thouless pumping and topology, Nat. Rev. Phys. **5**, 87 (2023).
[9] S.-C. Zhang and J. Hu, A Four-Dimensional Generalization of the Quantum Hall Effect, Science **294**, 823 (2001).
[10] Y. E. Kraus, Z. Ringel, and O. Zilberberg, Four-Dimensional Quantum Hall Effect in a Two-Dimensional Quasicrystal, Phys. Rev. Lett. **111**, 226401 (2013).
[11] W. A. Benalcazar, B. A. Bernevig, and T. L. Hughes, Quantized electric multipole insulators, Science **357**, 61 (2017).
[12] W. A. Benalcazar, B. A. Bernevig, and T. L. Hughes, Electric multipole moments, topological multipole moment pumping, and chiral hinge states in crystalline insulators, Phys. Rev. B **96**, 245115 (2017).
[13] Z. Song, Z. Fang, and C. Fang, (d-2)-Dimensional Edge States of Rotation Symmetry Protected Topological States, Phys. Rev. Lett. **119**, 246402 (2017).
[14] F. Schindler, A. M. Cook, M. G. Vergniory, Z. Wang, S. S. P. Parkin, B. A. Bernevig, and T. Neupert, Higher-order topological insulators, Sci. Adv. **4**, eaat0346 (2018).
[15] F. Schindler, Z. Wang, M. G. Vergniory, A. M. Cook, A. Murani, S. Sengupta, A. Y. Kasumov, R. Deblock, S. Jeon, I. Drozdov, H. Bouchiat, S. Guéron, A. Yazdani, B. A. Bernevig, and T. Neupert, Higher-order topology in bismuth, Nat. Phys. **14**, 918 (2018).
[16] B. Kang, W. Lee, and G. Y. Cho, Many-Body Invariants for Chern and Chiral Hinge Insulators, Phys. Rev. Lett. **126**,

016402 (2021).

[17] J. F. Wienand, F. Horn, M. Aidelsburger, J. Bibo, and F. Grusdt, Thouless Pumps and Bulk-Boundary Correspondence in Higher-Order Symmetry-Protected Topological Phases, Phys. Rev. Lett. **128**, 246602 (2022).

[18] I. Petrides and O. Zilberberg, Higher-order topological insulators, topological pumps and the quantum Hall effect in high dimensions, Phys. Rev. Res. **2**, 022049R (2020).

[19] B. Y. Xie, O. B. You, and S. Zhang, Photonic topological pump between chiral disclination states, Phys. Rev. A **106**, L021502 (2022).

[20] B. L. Wu, A. M. Guo, Z. Q. Zhang, and H. Jiang, Quantized charge-pumping in higher-order topological insulators, Phys. Rev. B **106**, 165401 (2022).

[21] Y. P. Wu, L. Z. Tang, G. Q. Zhang, and D. W. Zhang, Quantized topological Anderson-Thouless pump, Phys. Rev. A **106**, L051301 (2022).

[22] Y. E. Kraus, Y. Lahini, Z. Ringel, M. Verbin, and O. Zilberberg, Topological States and Adiabatic Pumping in Quasicrystals, Phys. Rev. Lett. **109**, 106402 (2012).

[23] S. Nakajima, T. Tomita, S. Taie, T. Ichinose, H. Ozawa, L. Wang, M. Troyer, and Y. Takahashi, Topological Thouless pumping of ultracold fermions, Nat. Phys. **12**, 296 (2016).

[24] M. Lohse, C. Schweizer, O. Zilberberg, M. Aidelsburger, and I. Bloch, A Thouless quantum pump with ultracold bosonic atoms in an optical superlattice, Nat. Phys. **12**, 350 (2016).

[25] O. Zilberberg, S. Huang, J. Guglielmon, M. Wang, K. P. Chen, Y. E. Kraus, and M. C. Rechtsman, Photonic topological boundary pumping as a probe of 4D quantum Hall physics, Nature **553**, 59 (2018).

[26] M. Lohse, C. Schweizer, H. M. Price, O. Zilberberg, and I. Bloch, Exploring 4D quantum Hall physics with a 2D topological charge pump, Nature **553**, 55 (2018).

[27] W. Chen, E. Prodan, and C. Prodan, Experimental Demonstration of Dynamic Topological Pumping across Incommensurate Bilayered Acoustic Metamaterials, Phys. Rev. Lett. **125**, 224301 (2020).

[28] I. H. Grinberg, M. Lin, C. Harris, W. A. Benalcazar, C. W. Peterson, T. L. Hughes, and G. Bahl, Robust temporal pumping in a magneto-mechanical topological insulator, Nat. Commun. **11**, 974 (2020).

[29] M. I. N. Rosa, R. K. Pal, J. R. F. Arruda, and M. Ruzzene, Edge States and Topological Pumping in Spatially Modulated Elastic Lattices, Phys. Rev. Lett. **123**, 034301 (2019).

[30] H. Chen, H. Zhang, Q. Wu, Y. Huang, H. Nguyen, E. Prodan, X. Zhou, and G. Huang, Creating synthetic spaces for higher-order topological sound transport, Nat. Commun. **12**, 5028 (2021).

[31] W. A. Benalcazar, J. Noh, M. Wang, S. Huang, K. P. Chen, and M. C. Rechtsman, Higher-order topological pumping and its observation in photonic lattices, Phys. Rev. B **105**, 195129 (2022).

[32] S. Y. Wang, Z. Hu, Q. Wu, H. Chen, E. Prodan, R. Zhu, and G. L. Huang, Smart patterning for topological pumping of elastic surface waves, Sci. Adv. **9**, eadh431 (2023).

[33] C. Y. Hou, C. Chamon, and C. Mudry, Electron fractionalization in two-dimensional graphenelike structures, Phys. Rev. Lett. **98**, 186809 (2007).

[34] J. Noh, W. A. Benalcazar, S. Huang, M. J. Collins, K. P. Chen, T. L. Hughes, and M. C. Rechtsman, Topological protection of photonic mid-gap defect modes, Nat. Photonics **12**, 408 (2018).

[35] W. A. Benalcazar, T. Li, and T. L. Hughes, Quantization of fractional corner charge in Cn-symmetric higher-order topological crystalline insulators, Phys. Rev. B **99**, 245151 (2019).

[36] E. Lee, A. Furusaki, and B. Yang, Fractional charge bound to a vortex in two-dimensional topological crystalline insulators, Phys. Rev. B **101**, 241109 (2020).

[37] H. Qiu, M. Xiao, F. Zhang, and C. Qiu, Higher-Order Dirac Sonic Crystals, Phys. Rev. Lett. **127**, 146601 (2021).

[38] W. Yao, S. A. Yang, and Q. Niu, Edge states in graphene: from gapped flat-band to gapless chiral modes, Phys. Rev. Lett. **102**, 96801 (2009).

[39] C.-K. Chiu, J. C. Y. Teo, A. P. Schnyder, and S. Ryu, Classification of topological quantum matter with symmetries, Rev. Mod. Phys. **88**, 035005 (2016).

[40] Y. Wang, Y. Ke, Y. Chang, Y. Lu, J. Gao, C. Lee, and X. Jin, Constructing higher-order topological states in higher dimensions, Phys. Rev. B **104**, 224303 (2021).

[41] Z. Chen, W. Zhu, Y. Tan, L. Wang, and G. Ma, Acoustic Realization of a Four-Dimensional Higher-Order Chern Insulator and Boundary-Modes Engineering, Phys. Rev. X **11**, 11016 (2021).

[42] H. Xue et al, Observation of an acoustic octupole topological insulator, Nat. Commun. **11**, 2442 (2020).

[43] X. Ni, M. Li, M. Weiner, A. Alù, and A. B. Khanikaev, Demonstration of a quantized acoustic octupole topological insulator, Nat. Commun. **11**, 2108 (2020).

[44] Y. Qi, C. Qiu, M. Xiao, H. He, M. Ke, and Z. Liu, Acoustic realization of quadrupole topological insulators, Phys. Rev. Lett. **124**, 206601 (2020).

[45] T. Li, et al. Acoustic Möbius insulators from projective symmetry, Phys. Rev. Lett. **128**, 116803 (2022).